\documentclass[prl,twocolumn,superscriptaddress,showpacs,amsmath,amssymb]{revtex4-1}
\usepackage[dvips]{graphicx}
\usepackage[dvips]{color}
\usepackage{subfigure}
\usepackage{natbib}
\usepackage{bm}
\usepackage{siunitx}
\def\U#1{{\rm #1}}

\begin{document}
\title{
Frequency-multiplexed photon pairs over 1000 modes from a quadratic nonlinear
optical waveguide resonator with a singly-resonant configuration
}
\author{Rikizo Ikuta}
\affiliation{Graduate School of Engineering Science, Osaka University,
  Toyonaka, Osaka 560-8531, Japan}
\affiliation{
Quantum Information and Quantum Biology Division, 
Institute for Open and Transdisciplinary Research Initiatives,
Osaka University, Osaka 560-8531, Japan}
\author{Ryoya Tani}
\affiliation{Graduate School of Engineering Science, Osaka University,
  Toyonaka, Osaka 560-8531, Japan}
\author{Masahiro Ishizaki}
\affiliation{Graduate School of Engineering Science, Osaka University,
Toyonaka, Osaka 560-8531, Japan}
\author{Shigehito~Miki}
\affiliation{Advanced ICT Research Institute, 
National Institute of Information and Communications Technology (NICT),
Kobe 651-2492, Japan}
\affiliation{
Graduate School of Engineering Faculty of Engineering, Kobe University, 
Kobe 657-0013, Japan}
\author{Masahiro~Yabuno}
\affiliation{Advanced ICT Research Institute, 
National Institute of Information and Communications Technology (NICT),
Kobe 651-2492, Japan}
\author{Hirotaka~Terai}
\affiliation{Advanced ICT Research Institute, 
National Institute of Information and Communications Technology (NICT),
Kobe 651-2492, Japan}
\author{Nobuyuki Imoto}
\affiliation{
Quantum Information and Quantum Biology Division, 
Institute for Open and Transdisciplinary Research Initiatives,
Osaka University, Osaka 560-8531, Japan}
\author{Takashi Yamamoto}
\affiliation{Graduate School of Engineering Science, Osaka University,
Toyonaka, Osaka 560-8531, Japan}
\affiliation{
Quantum Information and Quantum Biology Division, 
Institute for Open and Transdisciplinary Research Initiatives,
Osaka University, Osaka 560-8531, Japan}

\begin{abstract}
We demonstrate a frequency multiplexed photon pair generation 
based on a quadratic nonlinear optical waveguide 
inside a cavity which confines only signal photons
without confining idler photons and the pump light. 
We monolithically constructed the photon pair generator 
by a periodically-poled lithium niobate~(PPLN) waveguide 
with a high reflective coating for the signal photons around 1600~nm 
and with anti-refrective coatings for the idler photons around 1520~nm 
and the pump light at 780~nm at the end faces of the PPLN waveguide. 
We observed a comb-like photon pair generation 
with a mode spacing of the free spectral range of the cavity. 
Unlike the conventional multiple resonant photon pair generation experiments, 
the photon pair generation were incessant within a range of $80$~nm 
without missing teeth due to a mismatch of the energy conservation 
and the cavity resonance condition of the photons, 
resulting in over 1000-mode frequency multiplexed photon pairs in this range. 
\end{abstract}
\maketitle

Photon pairs produced by spontaneous parametric down conversion~(SPDC)
or spontaneous four wave mixing 
are commonly used as a resource of single photons and entangled photon pairs 
in photonic quantum information experiments. 
Unfortunately, the photon pairs include not only the genuine single photon pair 
but also multiple photon pair emission, 
which degrades the quality of the single-photon-based experiments. 
Typically, suppression of the multiple photon emission is achieved 
by setting a single photon emission probability to a value much smaller than unity, 
while it makes the amount of the vacuum large 
and the success probability of the protocol small.
To boost the success probability 
without increasing the photon emission rate per mode, 
parallel processing of the protocol~\cite{Caspani2017,Joshi2018} 
or use of a larger Hilbert space~\cite{Xie2015,Kues2017} is effective. 
For this purpose, 
frequency multiplexed photon pair generation has been actively studied.
Such a photon pair generator can be realized 
by using an optical parametric oscillator~(OPO) far below threshold~\cite{Lu2000}; 
A nonlinear optical medium is installed in an optical cavity for confining the photons, 
and a sufficiently weak pump light for the photon pair generation is used. 
Typical nonlinear optical media have a wide bandwidth for photon pair generation, 
whereas the optical cavity suppresses the photon generation in non-resonant modes. 
As a result, photon pairs with clear frequency mode separation are produced. 
So far, a lot of experiments of the photon pair generation 
with various cavity configurations have been performed~\cite{Caspani2017,Kues2019} 
by the quadratic nonlinerity with an external cavity~\cite{Wang2004,Kroh2017}, 
a quadratic nonlinearity-based microring resonator~\cite{Guo2017}, 
and Kerr nonlinearity-based microring resonators~\cite{Reimer2014,Reimer2016,Mazeas2016,Fujiwara2017,Kues2017,Jaramillo2017,Imany2018}. 
However, to the best of our knowledge, 
all of the previous demonstrations for frequency multiplexed photon pair generation 
have used a doubly-resonant OPO which confines both the signal and the idler photons, 
or a triply-resonant OPO which additionally confines the pump light. 
While the photon pair generation 
by using a singly-resonant~(SR) OPO which confines only one half of the photon pair 
has been demonstrated in Ref.~\cite{Scholz2007}, 
the experiment focused on the narrow-band single photon generation 
as in the case of the photon pair generation by the doubly-resonant OPO~\cite{Bao2008,Scholz2009,Pomarico2009,Yang2009,Pomarico2012,Chuu2012,Fekete2013,Luo2015,Niizeki2018}. 

In this paper, we demonstrate a frequency multiplexed photon pair generation 
based on a quadratic nonlinear optical medium with the SR OPO configuration. 
In the experiment, 
we used a periodically-poled lithium niobate~(PPLN) waveguide as 
the nonlinear optical medium which is widely used 
in the quantum information experiments~\cite{Alibart2016,Sharapova2017}
and has a possibility to integrate into an on-chip photonic circuit~\cite{Fortsch2013,Jin2014,Vergyris2016,Sansoni2017,Sharapova2017}. 
In our experiment,
a pair of photons is generated into a long and a short wavelength comb-like mode. 
For convenience, 
we call the long one and the short one as signal and idler photon, respectively. 
Our cavity is designed for confining only the signal photons 
in a monolithically-integrated quadratic nonlinear waveguide resonator~(QNWR) 
which we call PPLN-QNWR; 
dielectric multilayers are directly formed at the end faces of the PPLN waveguide.
The photon pair generation with the SR configuration has several features as follows. 
The photon pair generator does not need a severe stabilization 
of the pump laser frequency to the cavity. 
In addition, in spite of no cavity configuration for the idler photons, 
not only the signal photons but also the idler photons have comb-like spectrum  
as if both of them are confined in the cavity. 
Different from the photon pairs by the multiple resonant OPO, 
the spectra in SR configuration do not have missing teeth
within specific spectral intervals 
derived from a mismatch of the energy conservation 
and the cavity resonance conditions of the photons known as
the cluster effect~\cite{Pomarico2009,Pomarico2012,Chuu2012,Fekete2013,Luo2015}. 
As a result, a wide-band frequency multiplexed photon pair generation can be possible. 
When we regard the device as a single photon generator in the idler mode 
heralded by the signal photon of the photon pair, 
the idler photon does not suffer from the cavity loss 
and thus the narrow spectral photon can be efficiently prepared. 
The heralded idler photon has an exponential rising waveform 
as the time reversed waveform of the heralding signal photon confined in the cavity.
  This property is expected to achieve a high mode matching
  to an atomic resonance~\cite{Zhang2012,Srivathsan2014} 
  like the narrow-band photons
  via spontaneous four-wave mixing in atoms~\cite{Du2008,Zhao2014} 
  and optical circuits employing cavity systems~\cite{Bader2013,Liu2014}.

\begin{figure}
 \begin{center}
    \scalebox{1.2}{\includegraphics{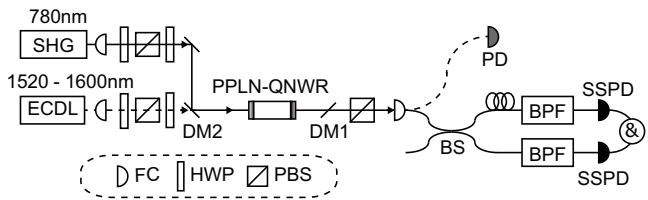}}
    \caption{Our experimental setup for photon pair generation.
      DM2 reflects the 780-nm light and transmits the telecom photons.
      FC is a fiber coupler. 
 }
 \label{fig:setup}
 \end{center}
\end{figure}
The experimental setup is shown in Fig.~\ref{fig:setup}. 
A continuous wave pump light at 780~nm for SPDC is 
frequency stabilized to a resonance of rubidium atoms~\cite{Tsujimoto:17}. 
By using a pair of half waveplates~(HWP) sandwiching a polarization beamsplitter~(PBS), 
the power of the pump light coupled to the waveguide
is set to 500~$\mu$W to 2~mW, 
and its polarization is set to vertical~(V) polarization. 
After that,
the pump light is focused on the PPLN-QNWR for SPDC
with the coupling efficiency of 0.9. 

\begin{figure}[t]
 \begin{center}
   \scalebox{1}{\includegraphics{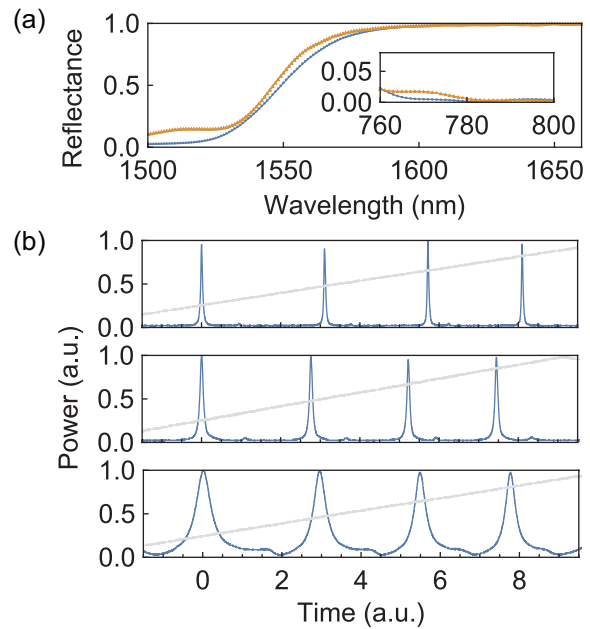}}
   \caption{
     (a) Reflectances of the coating at the surfaces of 
     lithium niobate samples coated 
     at the same batches for the PPLN waveguide.
     The inset shows the reflectances for around the pump wavelength.
     The orange and the blue curves are for 
       the back and the front ends of the PPLN, respectively. 
     (b)
     Observed power of the light passing through the the PPLN-QNWR
     for 1600-nm, 1580-nm and 1560-nm light from the top.
     The gray lines describe voltages applied to the ECDL 
     for the wavelength sweep. 
     Each peak is fitted by a function 
      $A((f-f_{0})^2+(\gamma_f/2)^2)^{-1}+d$
      with fitting parameters $A$, $f_0$, $\gamma_f$ and $d$. 
 }
 \label{fig:peaks}
 \end{center}
\end{figure}

The PPLN waveguide used in our experiment is a Zinc-doped lithium niobate 
as a core and lithium tantalite as a clad. The length is 20~mm. 
The periodically-poling period is \SI{18.090}{nm},
and the waveguide satisfies 
the type-0 quasi-phase-matching~(the polarization of the relevant three light is V)
of the second harmonic generation of 1560-nm light at 50~$^\circ$C. 
Both ends of the PPLN are flat polished for Fabry-P\'{e}rot cavity structure, 
and coated by the dielectric multilayers 
for a high reflective coating around 1600~nm and 
anti-reflective coatings around 1520~nm and 780~nm, 
which are shown in Fig.~\ref{fig:peaks}~(a). 
The signal and the idler photons generated at the PPLN-QNWR are separated 
from the pump light by a dichroic mirror~(DM1), 
and then the V-polarized photons are coupled to a single mode fiber~(SMF). 
The photons are divided into two paths by a fiber-based half beamsplitter~(BS). 
Both photons pass through bandpass filters~(BPF) with their minimum bandwidth of 0.03~nm, 
and finally they are detected 
by superconducting single photon detectors~(SSPDs)~\cite{Miki2017}. 
The electrical signals from the two detectors are connected 
to the time-to-digital converter~(TDC), 
and the coincidence counts with timestamps are recorded. 

\begin{table}[t]
\begin{center}
\begin{tabular}
  {|c|c|c|c|}
  \hline
  Wavelength & FWHM $\gamma_f$ & $Q$ factor & Finesse \\ \hline\hline
  1600~nm & 60~MHz & $3.2\times 10^6$ & 59 \\ \hline
  1580~nm  & 116~MHz & $1.6\times 10^6$ & 29 \\ \hline
  1560~nm & 470~MHz & $0.4\times 10^6$ & 7 \\ \hline
\end{tabular}
 \caption{
   The estimated FWHM $\gamma_f$ of the resonant peaks. 
 \label{tbl:fwhm}}
 \end{center}
\end{table}
Before the photon-pair generation, 
we first measured the optical response of the PPLN-QNWR. 
We turn off the pump laser at 780~nm, 
and we observe the transmission spectra of a telecom light 
from an external cavity diode laser~(ECDL)
with scanning the frequency.
The telecom light output from the PPLN-QNWR is 
coupled to the SMF, and is detected by a photo detector~(PD) 
followed by an oscilloscope. 
Examples of the observed resonant peaks are shown in Fig.~\ref{fig:peaks}~(b). 
Each peak is fitted by using a Lorentzian with a constant noise level. 
By borrowing the free spectral range~(FSR) $\Delta_f=3.5$~GHz of the cavity 
from a previous reported value by using a 20-mm-long PPLN waveguide 
with a resonant structure for telecom light~\cite{Ikuta2018-2}, 
we showed a full width at the half maximum~(FWHM) denoted by $\gamma_f$ 
in Table~\ref{tbl:fwhm}. 

\begin{figure}[t]
 \begin{center}
   \scalebox{1}{\includegraphics{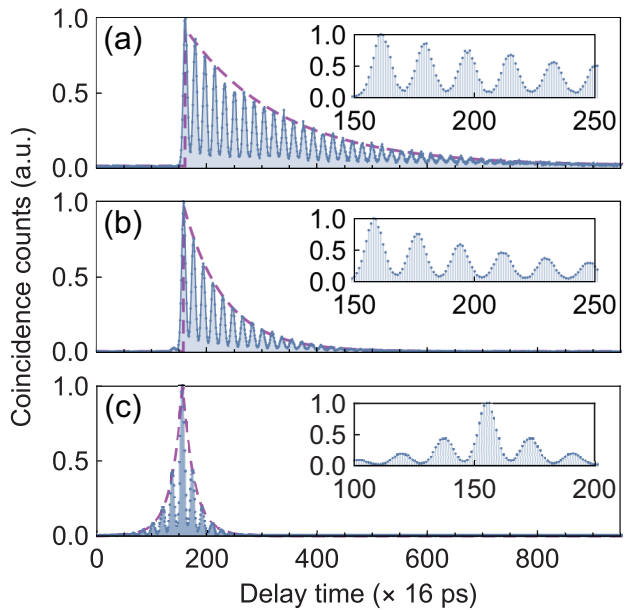}}
    \caption{
      Observed coincidence counts with the filter bandwidths of 1~nm. 
      The wavelengths of the signal and the idler photons are 
      (a)
      \SI{1600}{nm} \& \SI{1522}{nm} and
      (b)
      \SI{1580}{nm} \& \SI{1540}{nm}. 
      (c)
      The case of the degenerated photon pair around \SI{1560}{nm}. 
      The non zero values of the minimum of the fringes 
      is caused by the timing jitter about 80~ps of the SSPDs. 
    }
 \label{fig:oscillation}
 \end{center}
\end{figure}
From the above experimental result,
the photon pairs in the SR configuration are expected to be generated 
with their mode spacing corresponding to FSR of the cavity. 
To see this, 
we measured the beat signals of the frequency separated photons 
by the coincidence measurement.
We set the bandwidths of the BPFs to 1~nm for collecting 
the several separated frequency modes for both the signal and the idler photons.
The electrical signals from the detectors for the idler and the signal photons 
are used as a start and a stop of the TDC, respectively. 
In this setup, 
the beat effect among the frequencies of the detected photons should be observed 
as an oscillation of the coincidence counts between the signal
and the idler photons~\cite{Scholz2007,Herzog2008}.
In Fig.~\ref{fig:oscillation} from the top, We show 
the experimental results for signal and idler pairs in SR configuration 
around (\SI{1600}{nm} \& \SI{1522}{nm}) and (\SI{1580}{nm} \& \SI{1540}{nm}), 
and as a reference, the degenerated photon pair around 1560~nm 
in doubly-resonant configuration. 
From the figure, we clearly observed the oscillation of the coincidence, 
which indicates that
the photon pairs are in a superposition of
different frequency modes with a mode spacing. 
By picking up the peaks of the observed coincidence, 
we calculated the periodic time of the peaks 
about the photon pairs
around (\SI{1600}{nm} \& \SI{1522}{nm}), (\SI{1580}{nm} \& \SI{1540}{nm})
and (\SI{1560}{nm} \& \SI{1560}{nm}). 
The estimated time periods are \SI{286}{ps}, \SI{286}{ps} and \SI{282}{ps}, respectively, 
that correspond to the FSR $\Delta_f=3.5$~GHz of the cavity.
This shows that the photon pairs 
were surely produced with a mode spacing of the FSR.

In Fig.~\ref{fig:oscillation}, 
we see not only the oscillation but also the time decay of the coincidence counts. 
In the case of the doubly-resonant photon pair around 1560~nm,
the symmetric waveform in the time domain was observed
as in the previous experiments~\cite{Scholz2009,Fekete2013}. 
On the other hand, in the cases of
the photon pairs in SR configuration  
around (\SI{1600}{nm} \& \SI{1522}{nm}) and (\SI{1580}{nm} \& \SI{1540}{nm}), 
single-sided exponential time decays were observed. 
The best fit of the envelopes of the curves to the Lorentzian 
$A \exp(-\gamma (t-\tau_0)/2)+d$ for $t\geq \tau_0$ and $d$ for $t<\tau_0$ 
with fitting parameters $A$, $\gamma$ and $d$ shows that 
the FWHMs $\gamma/(2\pi)$ of the Lorentzian
in the frequency domain are 114~MHz and 270~MHz, respectively. 
Here $\tau_0$ is the time when the maximum of the coincidence counts was obtained.
The values are almost twice as large as 
the FWHMs $\gamma_f$ estimated by using the classical laser light in Table~\ref{tbl:fwhm}. 

\begin{figure}[t]
 \begin{center}
   \scalebox{1}{\includegraphics{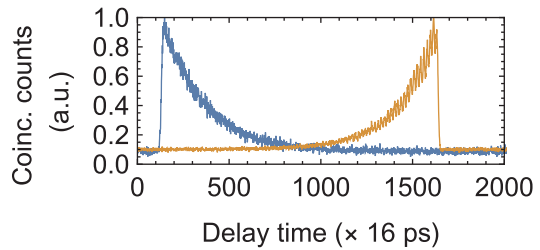}}
    \caption{
      The coincidence counts with exponential decay~(rising)
      described by blue~(orange) show 
      the signal~(idler) photon counts heralded by the idler~(signal) photons
      of the photon pair. 
      The measurement was performed with the bandwidths 0.03~nm of the BPFs. 
    }
 \label{fig:timerev}
 \end{center}
\end{figure}
Next we set the bandwidth of the BPFs for the signal and idler photons 
to 0.03~nm corresponds to 3.7~GHz. 
These filters severely limit the frequency modes to a single mode 
for both signal and idler. 
We set the center wavelengths of the BPFs to 1600~nm and 1522~nm. 
The observed coincidence counts are shown in Fig.~\ref{fig:timerev}. 
Due to the single frequency mode filtering, 
the exponential time decay without oscillation was observed. 
By swapping the role of the BPFs, 
the coincidence counts with the time reversed shape was obtained.
This result can be interpreted that 
the photon with the exponential rising shape is prepared in the idler mode 
heralded by the signal photon detection. 

\begin{figure}[t]
 \begin{center}
    \scalebox{1}{\includegraphics{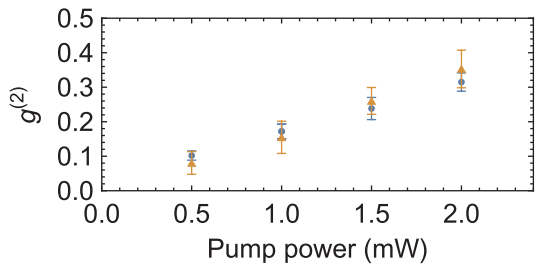}}
    \caption{
      $g^{(2)}$ vs pump power.
      The blue circles are about signal photons heralded by idle photons,
      and the orange triangles are about idler photons heralded by signal photons. 
 }
 \label{fig:g2}
 \end{center}
\end{figure}
For the above wavelengh setting, 
we measured the second-order intensity correlation function $g^{(2)}$ 
of the signal~(idler) photon heralded by the idler~(signal) photon. 
The measurement was performed by splitting photons 
into two spatial paths after one of the BPFs in Fig.~\ref{fig:setup}.
The coincidence time window was set to 
9~ns~(corresponding to $\sim$ 560 bin in Fig.~\ref{fig:timerev}),
in which a large portion of the wave packet was included. 
The experimental result is shown in Fig.~\ref{fig:g2}. 
As is the case with the conventional SPDC process, 
the functions $g^{(2)}$ of the heralded signal and idler photons take the same value, 
and are proportional to the pump power in this power range.

\begin{figure}[t]
 \begin{center}
   \scalebox{1}{\includegraphics{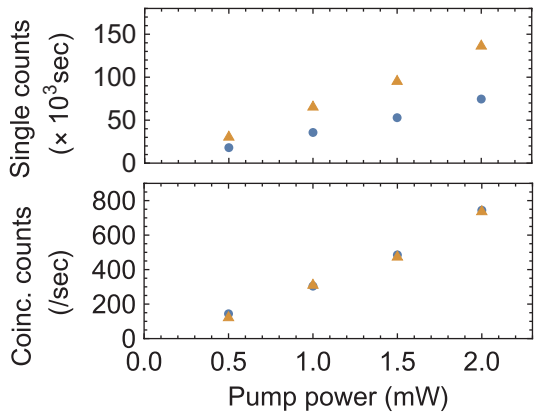}}
   \caption{
     Pump dependencies of 
     single counts of signal photons~(blue circles)
     and idler photons~(orange triangles),
     and 
     coincidence counts 
     when signal~(idler) photons 
     are used as start of TDC
     which are plotted by blue circles~(orange triangles). 
 }
 \label{fig:counts}
 \end{center}
\end{figure}
The single and coincidence counts 
are also proportional to the pump power as shown in Fig.~\ref{fig:counts}. 
The ratio of the single counts of the the signal photons 
to that of the idler photons is about 0.54. 
The coupling efficiencies of the signal and the idler photons 
were estimated to be almost the same by using the classical laser light, 
and thus 
the difference of the single counts reflects our cavity design 
such that the reflectance at the both ends of the PPLN-QNWR for 1600~nm 
is the same as shown in Fig.~\ref{fig:peaks}~(a). 
If a proper high reflective coating is formed at the front end of the PPLN, 
all signal photons will come out from the back end of the PPLN. 
From the ratio of the coincidence counts 
to the single counts of the signal~(idler) photon, 
we estimated the transmittance of the optical circuit for the idler~(signal) photon. 
The estimated overall transmittances for the signal and the idler photons
just before the photon detectors are 0.009 and 0.013, respectively, 
with an assumption of the quantum efficiency 0.6 of the SSPDs.
The values agree well with those measured by using classical laser light,
in which the fiber coupling and frequency filtering efficiencies were 0.3 and 0.05. 
As a result,
the intrinsic photon pair rate emitted from the back end of the PPLN-QNWR 
within the single resonant pair for 1600~nm and 1522~nm 
is estimated as $4\times 10^6$~$\U{pairs/(s\cdot mW)}$.

\begin{figure}[t]
 \begin{center}
    \scalebox{1}{\includegraphics{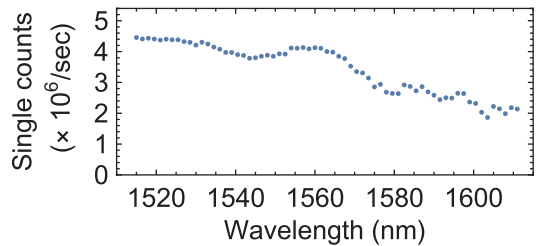}}
    \caption{
      Observed single counts.
      The bandwidth of the BPF is set to 3~nm. 
 }
 \label{fig:single}
 \end{center}
\end{figure}
Finally, in order to see no cluster effect in the photon pairs
produced by the PPLN-QNWR with SR configuration, 
we removed the BS in Fig.~\ref{fig:setup}, and 
measured the single photon counts by scanning the center wavelengths of the BPF. 
The bandwidth of the BPF is set to 3~nm and the pump power is set to 500~$\mu$W. 
The experimental result is shown in Fig.~\ref{fig:single}. 
Different from the cases of the SPDC based on the doubly-resonant cavity~\cite{Pomarico2009,Pomarico2012}, 
the photon pairs were not suppressed over all the spectral range
from \SI{1520}{nm} to \SI{1600}{nm}. 
The expected number of the photon pair modes 
with the mode spacing of the FSR is about 1400 in this range.  
We note that the single photon counts for longer wavelengths are smaller, 
because of our asymmetric cavity design for the signal and the idler photons 
as is the case with Fig.~\ref{fig:counts}~(a). 

In conclusion, 
we have demonstrated the frequency multiplexed photon pair generation 
based on the PPLN waveguide as the nonlinear optical crystal 
with the SR OPO configuration in which only signal photons are confined. 
As is the cases of the photon pair generation based on the multiple resonant OPO, 
we observed the oscillation of the coincidence counts
between the signal and the idler photons. 
This shows that the frequency modes of the photon pairs 
are well separated with the mode spacing corresponding to the FSR of the cavity 
as if both photons are confined in the cavity. 
The comb-like spectra of the photons have no missing teeth 
due to the elimination of the cluster effect which is seen in the photon pairs
based on the PPLN crystal in the doubly-resonant OPO. 
By using the signal photon confined in the cavity as the heralding photon, 
the heralded idler photons are efficiently extracted 
with the exponential rising waveform as the time reversed shape 
of the heralding signal photons. 
We believe the wide-band frequency multiplexed photon pair generator 
based on the PPLN waveguide resonator will be useful 
in various kinds of photonic quantum information processing 
such as an efficient quantum key distribution~\cite{Aktas2016, Wengerowsky2018},
an efficient photon-matter interface with an atomic frequency comb~\cite{Afzelius2009,Saglamyurek2016,Alessandro2017}.
In addition, the cavity design with different finesses 
depending on the frequency will be a significant progress for manipulation of 
the temporal mode~\cite{Reddy2018} 
and frequency mode~\cite{Menicucci2008,Xie2015,Kobayashi2016,Lu2018,Joshi2018} 
with the use of frequency-domain optical elements~\cite{Ikuta2011,Lukens2017,Ikuta2018}. 

\begin{acknowledgements}
We thank Motoki Asano and Yoshiaki Tsujimoto for fruitful discussion. 
This work was supported 
by CREST, JST JPMJCR1671; MEXT/JSPS KAKENHI Grant Number 
JP16H02214
and JP18K13483; 
Asahi Glass Foundation, Murata Science Foundation, 
Fujikura Foundation, and 
Kayamori Foundation of Information Science Advancement. 
\end{acknowledgements}


\begin{thebibliography}{10}

\bibitem{Caspani2017}
L.~Caspani {\em et~al.},
\newblock Light: Science \& Applications {\bf 6}, e17100 (2017).

\bibitem{Joshi2018}
C.~Joshi, A.~Farsi, S.~Clemmen, S.~Ramelow, and A.~L. Gaeta,
\newblock Nature communications {\bf 9}, 847 (2018).

\bibitem{Xie2015}
Z.~Xie {\em et~al.},
\newblock Nature Photonics {\bf 9}, 536 (2015).

\bibitem{Kues2017}
M.~Kues {\em et~al.},
\newblock Nature {\bf 546}, 622 (2017).

\bibitem{Lu2000}
Y.~J. Lu and Z.~Y. Ou,
\newblock Phys. Rev. A {\bf 62}, 033804 (2000).

\bibitem{Kues2019}
M.~Kues {\em et~al.},
\newblock Nature Photonics {\bf 13}, 170 (2019).

\bibitem{Wang2004}
H.~Wang, T.~Horikiri, and T.~Kobayashi,
\newblock Phys. Rev. A {\bf 70}, 043804 (2004).

\bibitem{Kroh2017}
T.~Kroh, A.~Ahlrichs, B.~Sprenger, and O.~Benson,
\newblock Quantum Science and Technology {\bf 2}, 034007 (2017).

\bibitem{Guo2017}
X.~Guo {\em et~al.},
\newblock Light: Science \& Applications {\bf 6}, e16249 (2017).

\bibitem{Reimer2014}
C.~Reimer {\em et~al.},
\newblock Optics express {\bf 22}, 6535 (2014).

\bibitem{Reimer2016}
C.~Reimer {\em et~al.},
\newblock Science {\bf 351}, 1176 (2016).

\bibitem{Mazeas2016}
F.~Mazeas {\em et~al.},
\newblock Optics Express {\bf 24}, 28731 (2016).

\bibitem{Fujiwara2017}
M.~Fujiwara, R.~Wakabayashi, M.~Sasaki, and M.~Takeoka,
\newblock Optics express {\bf 25}, 3445 (2017).

\bibitem{Jaramillo2017}
J.~A. Jaramillo-Villegas {\em et~al.},
\newblock Optica {\bf 4}, 655 (2017).

\bibitem{Imany2018}
P.~Imany {\em et~al.},
\newblock Optics express {\bf 26}, 1825 (2018).

\bibitem{Scholz2007}
M.~Scholz, F.~Wolfgramm, U.~Herzog, and O.~Benson,
\newblock Applied Physics Letters {\bf 91}, 191104 (2007).

\bibitem{Bao2008}
X.-H. Bao {\em et~al.},
\newblock Phys. Rev. Lett. {\bf 101}, 190501 (2008).

\bibitem{Scholz2009}
M.~Scholz, L.~Koch, and O.~Benson,
\newblock Phys. Rev. Lett. {\bf 102}, 063603 (2009).

\bibitem{Pomarico2009}
E.~Pomarico {\em et~al.},
\newblock New Journal of Physics {\bf 11}, 113042 (2009).

\bibitem{Yang2009}
J.~Yang {\em et~al.},
\newblock Phys. Rev. A {\bf 80}, 042321 (2009).

\bibitem{Pomarico2012}
E.~Pomarico, B.~Sanguinetti, C.~I. Osorio, H.~Herrmann, and R.~T. Thew,
\newblock New Journal of Physics {\bf 14}, 033008 (2012).

\bibitem{Chuu2012}
C.-S. Chuu, G.~Y. Yin, and S.~E. Harris,
\newblock Applied Physics Letters {\bf 101}, 051108 (2012).

\bibitem{Fekete2013}
J.~Fekete, D.~Riel\"ander, M.~Cristiani, and H.~de~Riedmatten,
\newblock Phys. Rev. Lett. {\bf 110}, 220502 (2013).

\bibitem{Luo2015}
K.-H. Luo {\em et~al.},
\newblock New Journal of Physics {\bf 17}, 073039 (2015).

\bibitem{Niizeki2018}
K.~Niizeki {\em et~al.},
\newblock Applied Physics Express {\bf 11}, 042801 (2018).

\bibitem{Alibart2016}
O.~Alibart {\em et~al.},
\newblock Journal of Optics {\bf 18}, 104001 (2016).

\bibitem{Sharapova2017}
P.~Sharapova {\em et~al.},
\newblock New Journal of Physics {\bf 19}, 123009 (2017).

\bibitem{Fortsch2013}
M.~F{\"o}rtsch {\em et~al.},
\newblock Nature communications {\bf 4}, 1818 (2013).

\bibitem{Jin2014}
H.~Jin {\em et~al.},
\newblock Phys. Rev. Lett. {\bf 113}, 103601 (2014).

\bibitem{Vergyris2016}
P.~Vergyris {\em et~al.},
\newblock Scientific reports {\bf 6}, 35975 (2016).

\bibitem{Sansoni2017}
L.~Sansoni {\em et~al.},
\newblock npj Quantum Information {\bf 3}, 5 (2017).

\bibitem{Herzog2008}
U.~Herzog, M.~Scholz, and O.~Benson,
\newblock Phys. Rev. A {\bf 77}, 023826 (2008).

\bibitem{Jeronimo-Moreno2010}
Y.~Jeronimo-Moreno, S.~Rodriguez-Benavides, and A.~B. U'Ren,
\newblock Laser Physics {\bf 20}, 1221 (2010).

\bibitem{Bateman1954}
\newblock H.~Bateman{\em , Tables of integral transforms} Vol. 1, Sec. 3.2,
  Eqs.~(3) and (4) (McGraw-Hill Book Company, 1954).

\bibitem{Zhang2012}
S.~Zhang {\em et~al.},
\newblock Phys. Rev. Lett. {\bf 109}, 263601 (2012).

\bibitem{Srivathsan2014}
B.~Srivathsan, G.~K. Gulati, A.~Cer\`e, B.~Chng, and C.~Kurtsiefer,
\newblock Phys. Rev. Lett. {\bf 113}, 163601 (2014).

\bibitem{Du2008}
S.~Du, P.~Kolchin, C.~Belthangady, G.~Y. Yin, and S.~E. Harris,
\newblock Phys. Rev. Lett. {\bf 100}, 183603 (2008).

\bibitem{Zhao2014}
L.~Zhao {\em et~al.},
\newblock Optica {\bf 1}, 84 (2014).

\bibitem{Bader2013}
M.~Bader, S.~Heugel, A.~L. Chekhov, M.~Sondermann, and G.~Leuchs,
\newblock New Journal of Physics {\bf 15}, 123008 (2013).

\bibitem{Liu2014}
C.~Liu {\em et~al.},
\newblock Phys. Rev. Lett. {\bf 113}, 133601 (2014).

\bibitem{Tsujimoto:17}
Y.~Tsujimoto {\em et~al.},
\newblock Opt. Express {\bf 25}, 12069 (2017).

\bibitem{Miki2017}
S.~Miki, M.~Yabuno, T.~Yamashita, and H.~Terai,
\newblock Optics Express {\bf 25}, 6796 (2017).

\bibitem{Ikuta2018-2}
R.~Ikuta, M.~Asano, R.~Tani, T.~Yamamoto, and N.~Imoto,
\newblock Optics Express {\bf 26}, 15551 (2018).

\bibitem{Aktas2016}
D.~Aktas {\em et~al.},
\newblock Laser \& Photonics Reviews {\bf 10}, 451 (2016).

\bibitem{Wengerowsky2018}
S.~Wengerowsky, S.~K. Joshi, F.~Steinlechner, H.~H{\"u}bel, and R.~Ursin,
\newblock Nature {\bf 564}, 225 (2018).

\bibitem{Afzelius2009}
M.~Afzelius, C.~Simon, H.~de~Riedmatten, and N.~Gisin,
\newblock Phys. Rev. A {\bf 79}, 052329 (2009).

\bibitem{Saglamyurek2016}
E.~Saglamyurek {\em et~al.},
\newblock Nature Communications {\bf 7}, 11202 (2016).

\bibitem{Alessandro2017}
A.~Seri {\em et~al.},
\newblock Phys. Rev. X {\bf 7}, 021028 (2017).

\bibitem{Reddy2018}
D.~V. Reddy and M.~G. Raymer,
\newblock Optics express {\bf 26}, 28091 (2018).

\bibitem{Menicucci2008}
N.~C. Menicucci, S.~T. Flammia, and O.~Pfister,
\newblock Phys. Rev. Lett. {\bf 101}, 130501 (2008).

\bibitem{Kobayashi2016}
T.~Kobayashi {\em et~al.},
\newblock Nature Photonics {\bf 10}, 441 (2016).

\bibitem{Lu2018}
H.-H. Lu {\em et~al.},
\newblock Optica {\bf 5}, 1455 (2018).

\bibitem{Ikuta2011}
R.~Ikuta {\em et~al.},
\newblock Nature Communications {\bf 2}, 1544 (2011).

\bibitem{Lukens2017}
J.~M. Lukens and P.~Lougovski,
\newblock Optica {\bf 4}, 8 (2017).

\bibitem{Ikuta2018}
R.~Ikuta {\em et~al.},
\newblock Nature Communications {\bf 9}, 1997 (2018).

\end{thebibliography}
\end{document}